\documentclass[12pt]{article}
\usepackage{amsmath,amssymb,graphicx,mathrsfs,slashed,setspace}
\usepackage{amsmath, amssymb, graphicx,setspace}
\usepackage{amsfonts}
\usepackage[utf8]{inputenc}
\usepackage[T1]{fontenc}
\usepackage{mathrsfs}
\usepackage[normalem]{ulem}
\newcommand{\be}{\begin{equation}}
\newcommand{\ee}{\end{equation}}
\newcommand{\bea}{\begin{eqnarray}}
\newcommand{\eea}{\end{eqnarray}}

\def\({\left(} \def\){\right)}

\usepackage{float}
\usepackage{amsmath}
\usepackage{amssymb}
\usepackage{graphicx}
\usepackage[letterpaper]{geometry}
\usepackage{setspace}
\usepackage{babel}
\usepackage[
  colorlinks=true,
  linkcolor=blue,
  urlcolor=blue,
  citecolor=green
]{hyperref}
\begin{document}
\renewcommand{\baselinestretch}{1.5}

	\title{
{{Resonance correction \\ to time-dependent finite-size effects \\ in Effective Field Theory of binary black hole inspiral} }}
	\author{\large Shani Avitan, Ram Brustein \\
		\vspace{-.5in}  \vbox{
			\begin{center}
				 Department of Physics, Ben-Gurion University,
						Beer-Sheva 84105, Israel\\\vspace{.05in}	
				\ \\  
				shavi@post.bgu.ac.il,\ ramyb@bgu.ac.il
			\end{center}
	}}
	\renewcommand{\baselinestretch}{1.35}\normalsize
	\date{}
	\maketitle
\begin{abstract}	
\noindent
General Relativity predicts that black holes lack an internal structure and that their tidal Love numbers vanish. However, if astrophysical black holes were instead ultra-compact objects (UCOs) with internal structure, their tidal Love numbers would no longer vanish. Then, during a binary inspiral, their tidal deformation could transfer orbital energy to the UCO, increasing the orbital period and leaving an observable imprint on the phase of the emitted gravitational waves. As we discussed previously, the effect is enhanced if one binary companion is resonantly excited. Here we use the point-particle effective field theory framework to calculate this enhancement. We consider circular orbits and focus on excitations of the internal electric quadrupole, modelling the UCO spectrum as a harmonic oscillator spectrum. However, our methods could be straightforwardly extended to more complicated situations. Integrating out the internal quadrupole induces a series of time-dependent finite-size operators with specific numerical coefficients, which we calculate. We find that the static electric quadrupole Love number $k_2$ is negative, and that all finite-size terms are proportional to $k_2$. We further show that resonant excitation induces a leading 2.5 Post-Newtonian term, compared with the 5 Post-Newtonian term in the time-independent case. Our results confirm the expectation that resonant excitation during a binary inspiral offers one of the best ways to detect whether astrophysical black holes can undergo tidal deformation and, if they do, learn about their internal excitation spectrum \cite{Avitan:2023txy}.
\end{abstract}
	\newpage
\renewcommand{\baselinestretch}{1.5}\normalsize
\section{Introduction}

General Relativity (GR) predicts that black holes (BHs) lack an internal structure, but is this prediction correct? A way to put the GR prediction to the test is to compare it to the prediction of an alternative description of astrophysical BHs as ultra-compact objects (UCOs), whose external geometry is identical to Schwarzschild or Kerr geometry. These are the so-called blackhole mimickers \cite{Bambi:2025wjx,Carballo-Rubio:2025fnc,Mathur:2026zml,Mathur:2024ify,Alho:2023mfc,Danielsson:2023onu,Liebling:2012fv,Bah:2022yji,Dima:2025tjz,Batic:2026zrz,HassanPuttasiddappa:2026kxm,Riojas:2026god,Maurya:2026lae,Mazur:2001fv,Brustein:2025yvf,Brustein:2026idb}.

An important additional motivation for introducing BH mimickers is to avoid the paradoxical issues associated with the  inevitable gravitational collapse expected to occur for compact objects beyond the Buchdahl bound \cite{Buchdahl:1959zz} and predicted by the celebrated singularity theorems \cite{Hawking:1970zqf}. Maximally negative radial pressure throughout the volume of the object can oppose gravity and circumvent the Buchdahl bound \cite{Brustein:2021lnr,Song:2026zvj,Andreasson:2025uir,Mazur:2015kia}.

Binary black holes systems provide opportunities to learn about BHs when the gravitational field is strong and possibly verify or refute the GR prediction. The data about their emitted gravitational waves (GW) is coming in at an increased rate and are expected to further improve in the future due to ongoing and planned GW observations \cite{LISA:2024hlh,LIGOScientific:2026wfs,Shoemaker:2026rfx,ET:2025xjr,Du:2025xdq,Luo:2025sos,Komori:2025pjx,Arduini:2026vxb}.

If the true state of the interior of astrophysical BHs were not accurately described by GR, but rather by UCOs, which do possess an internal structure, they could be excited and deformed when an external gravitational field acts on them, as is the case during a binary inspiral.  The deformations of the mass multipole moments are parametrized by tidal response coefficients, commonly referred to as the static Love numbers $k_{l}$, which characterize the linear relation between the induced multipole moments and the external tidal field.

During an inspiral, the source of energy absorbed by the UCO is orbital energy. Consequently, the orbital energy decreases, and the orbital frequency increases relative to the Keplerian value. This increase in orbital frequency results in a phase shift of the emitted GW. With improved sensitivity and a broader frequency range, GW observatories could hopefully detect this phase shift.

An important result is that, in the GR description of BHs in four spacetime dimensions, the static tidal Love numbers exactly vanish \cite{Binnington:2009bb,Damour:2009vw,Hui:2021vcv,Charalambous:2021mea,Kol:2011vg,Ivanov:2022qqt,Chia:2020yla}.
Moreover, recent studies show that the static tidal response coefficients
vanish beyond the linear order in the tidal potential \cite{Hadad:2024lsf,Riva:2023rcm,Iteanu:2024dvx,Combaluzier-Szteinsznaider:2024sgb,Gounis:2024hcm,Kehagias:2024rtz}.

Among the static Love numbers $k_l$, the electric quadrupolar Love number $k_{2}$ is expected to be the largest in a broad class of UCO models. This makes $k_2$ a key observable for testing GR. However, in the adiabatic part of the inspiral, tidal-response effects introduce only small corrections to the GW waveform at the fifth Post-Newtonian (PN) order \cite{Sennett:2017etc,Hotokezaka:2016bzh,Cardoso:2017cfl,Uchikata:2016qku,Cardoso:2019rvt}, making the detection of such an effect challenging.

Time-dependent tidal interactions have been studied perturbatively by expanding the linear tidal response functions for small, finite orbital frequencies, with the corresponding frequency-dependent coefficients commonly referred to as dynamical Love numbers \cite{Ivanov:2022hlo,Chakraborty:2025wvs,Porto:2016pyg}. These coefficients are not fixed by the static Love numbers and have been found to be nonvanishing. For Schwarzschild BHs, the leading conservative dynamical Love numbers arise quadratically in the tidal frequency and have been estimated to enter the GW phase at eighth PN order.

In an earlier work, we proposed the possibility of measuring $k_{2}$ through resonant tidal excitation, and argued that the tidal-induced correction can be significantly
enhanced relative to the adiabatic case, perhaps even within observational reach \cite{Avitan:2023txy}. Tidal resonant excitation requires that the frequency of an internal mode becomes equal to the orbital frequency $\Omega$, as the orbital frequency increases and the orbital radial distance decreases during the binary inspiral. Focusing on the electric quadrupole mode and denoting its characteristic frequency by $\omega_2$, the condition for a resonance to occur is
$
\omega_{2}\simeq 2\Omega.
$
This requires that the mode frequency $\omega_{2}$ is smaller than the natural frequency  $\omega_{2}\ll c/R_S$. As noted in \cite{Brustein:2020tpg}, this is to be expected for a UCO,  since the Tolman redshift factor at the surface of the UCO is large (rather than infinite, as in the case of a GR BH). This large redshift reduces the internal frequency from an external perspective.

In this paper, we use the point-particle (PP) effective field theory (EFT) framework at the orbital scale, focusing on effects related to the internal electric quadrupole excitation and modeling the UCO’s spectrum by that of a harmonic oscillator. Integrating out the internal quadrupole, using the exact Green function of the theory, induces a series of time-dependent finite-size (FS) operators with specific numerical coefficients, which we calculate. We find that the electric quadrupole Love number $k_2$ is negative, that all FS terms are proportional to $k_2$, and we verify that the resonant excitation induces a leading 2.5 PN term, compared with the 5 PN term in the time-independent case. Our results confirm the expectation that observing the effect of resonant excitation during a binary inspiral offers one of the best ways to detect whether astrophysical black holes can undergo tidal deformation and, if they do, learn about their internal excitation spectrum \cite{Avitan:2023txy}.

To discuss electric quadrupolar effects, the curved spacetime of the orbital motion is introduced through a quadrupolar metric correction to the flat background, which in turn couples to the induced quadrupole moment associated with the internal structure of the UCO. The tidal response is incorporated in the EFT by adding the so-called FS terms to the PP EFT. Such FS terms have been used to calculate the Love numbers for the static case of negligible orbital frequencies ($\Omega=0$)
\cite{Ivanov:2022hlo,Barbosa:2025uau,Goldberger:2020fot,Riva:2023rcm,Hui:2020xxx,Bhattacharyya:2025slf},
and in the context of absorption and tidal response also for small finite
frequencies \cite{DeLuca:2022tkm,Chakraborty:2025wvs,Charalambous:2021mea,Poisson:2020vap,Bhatt:2024yyz}.
Separating the background and induced quadrupole was discussed previously in \cite{Combaluzier-Szteinsznaider:2024sgb,Porto:2016pyg,Rai:2024lho,Saketh:2023bul,Cano:2025zyk}.
In a similar manner, we introduce a background quadrupole associated with the orbital motion of the binary and an internal quadrupolar degree of
freedom which can represent the $l=2$ mode of the internal structure
of a UCO.

The rest of the paper is organized as follows. In Section 2, we discuss generalities about the EFT and fix our conventions and notations. In Section 3, we briefly review the PP EFT. In Section 4, we discuss FS terms in the EFT. In Section 5, we discuss the properties of the model describing the time dependence within the EFT framework, and in Section 6 we discuss the limiting static case. In Section 7, we discuss perturbative time dependence, and in Section 8, we discuss the case of a resonant excitation of the internal modes by the orbital motion. In Section 9, we discuss the energy transfer from the orbit to the UCO and calculate the resulting phase shift of the orbital motion.  We end the paper with a brief conclusions and outlook section. Several technical discussions are relegated to five appendices,  Appendix A: introducing a simplified model of coupled harmonic oscillators for which the process of integrating out one degree of freedom can be performed explicitly, Appendix B: describing the calculation of the retarded Green function, Appendix C: performing the contraction of symmetric trace-free (STF) tensors, needed for the calculation of the FS terms and Appendix D: outlining the stationary phase approximation that we use to perform some integrals. Appendix E contains a discussion of time derivatives of the orbital frequency, showing that they are parametrically small.

\section{Effective Field Theory}

The PP EFT formalism provides a framework for describing the dynamics of a binary system that involves interactions on different, well-separated length scales  \cite{Donoghue:1995cz,Burgess:2003jk}. During a binary inspiral, the system radiates GW of wavelength $\lambda$, the orbital separation scale is $r$, and the Schwarzschild radius of the object is $R_{S}$. The ratios of the three relevant scales are
proportional to the small parameter $\dfrac{R_{S}}{r}\sim v^{2}$,
$\dfrac{r}{\lambda}\sim v$, $v$ being the orbital velocity. During the inspiral, the companions are moving relatively slow, remain far apart, and follow approximately Keplerian trajectories. The orbital velocity $v$ is parametrically small, $v^{2}\ll1$, so it is possible to use the PN expansion \cite{Poisson:1998ba,Blanchet:2013haa}.

It is convenient to start with a point-mass
moving in a curved space-time, whose mass is the reduced mass of the
binary companions. This starting point is most accurate in the early stages of the
Inspiral or in systems with a small mass ratio, such as extreme mass
ratio inspiral (EMRI) events, which LISA is expected to observe in
abundance \cite{LISA:2024hlh}.

For completeness and to establish our notation and conventions, we review the static limit, setting the orbital frequency $\Omega=0$. Next, we perform the perturbative expansion in $\Omega$. We find a connection between the tidal response coefficients to all orders in $\Omega$, resulting from their expression as an expansion of the Green function,  similar to the discussion in \cite{DeLuca:2024ufn}. Finally, we discuss the case in which the internal quadrupole mode frequency can be comparable
to the orbital frequency and discuss resonant excitation of the internal quadrupole mode. We are able to calculate precisely the coefficients of the FS terms in the action, allowing us to simplify, clarify, verify, and improve the results of our previous paper \cite{Avitan:2023txy} on tidal resonance.
We use natural units to describe gravitationally bound systems. In these units
Newton's constant $G=1$, the speed of light $c=1$,  mass and time have units of length, $m\sim[L]$, $t\sim [L]$. Our signature convention is $(-,+,+,+)$.

\section{Point mass in curved spacetime}

We discuss a binary system  during the inspiral
phase of a circular orbit. The companion masses are $M$ and $m$. We are mainly interested in the case  $M \gg m$, and define the mass ratio $q=\dfrac{m}{M}\ll 1$. We clarify the reason for our interest in this case later. We model the binary system using an effective PP description
with reduced mass $\mu=\dfrac{mM}{m+M}=\dfrac{q}{1+q}M \simeq q M$, coupled to
gravity. As explained in the introduction, our main focus will be on the finite-size effects related to the internal electric quadrupole, however, our methods can be straightforwardly applied to other internal excitations and to more general orbits.

The spacetime is expanded around Minkowski spacetime, with
the different contributions organized according to their physical origin.  The metric is decomposed as follows,
\begin{equation}
g_{\mu\nu}=\eta_{\mu\nu}+H^{BG}_{\mu\nu}+H^{I}_{\mu\nu},
\end{equation}
where $H^{BG}_{\mu\nu}$ corresponds to the background gravitational field
sourced by the orbital motion, and $H^{I}_{\mu\nu}$ corresponds to the part of
the metric sourced by the induced multipole moments associated with the internal structure of the compact objects. Indices are raised and lowered with $\eta_{\mu\nu}$.  Here we restrict the discussion to the case that both $H^{BG}_{\mu\nu}$ and $H^{I}_{\mu\nu}$ are of electric quadrupolar nature. The time-time component of the background contribution is given by
\begin{equation}
H^{BG}_{00}=Q^{BG}_{ij}x^{i}x^{j},
\end{equation}
where
\begin{equation}
Q^{BG}_{ij}=\left(\partial_{i}\partial_{j}U_{ext}\right)_{STF}=\frac{\mu}{r^{3}}\left(3n_{i}n_{j}-\delta_{ij}\right),\label{eq:BGquadrupolarSTF_def}
\end{equation}
with $r=\left|\vec{x}\right|$ being the orbital separation and
$n_{i}=\dfrac{x_{i}}{r}$. We use a coordinate system in which
$\vec{n}=\left(\cos\Omega\tau,\sin\Omega\tau,0\right)$, where $\Omega$
is the orbital angular velocity. The background quadrupole $Q^{BG}_{ij}$ is a symmetric
trace-free (STF) tensor scaling as $\left[L^{-2}\right]$. The
spatial components are parametrically smaller than the temporal components due to their higher-order
dependence on the orbital velocity $v$,
\begin{equation}
H^{BG}_{i0}\sim vH^{BG}_{00},\quad H^{BG}_{ij}\sim v^{2}H^{BG}_{00}
\end{equation}
and therefore can be neglected during the inspiral phase of the binary
evolution, if one considers the leading order in the PN expansion.

The induced quadrupole moment, defined as
\begin{equation}
Q^{I}_{ij}=\int d^{3}x\ \rho\left(x_{i}x_{j}-\frac{1}{3}\delta_{ij}r^{2}\right),
\end{equation}
with $\rho$ being the compact object's mass density, is
related to $H^{I}_{00}$,
\begin{equation}
H^{I}_{00}=C_{I}Q^{I}_{ij}x^{i}x^{j}.
\end{equation}
where the induced tensor $Q^{I}_{ij}$ is an STF tensor that scales as $\sim M R^{2}_{S}$, so  $Q^{I}_{ij}\sim\left[L^{3}\right]$.
Since $H^{I}$ is dimensionless, the coefficient $C_{I}$ scales as $\left[L^{-5}\right]$.

We aim to write an effective action for the orbital length scale $r$, therefore, we will choose $C_{I}\sim r^{-5}$. The parameter characterizing the smaller scale $R_S$ appears in the EFT action only after $Q^{I}_{ij}$ has been integrated out.

The action for a massive point particle moving in a curved spacetime can be written
as
\begin{equation}
S=S_{EH}+S_{pp},
\end{equation}
where the Einstein-Hilbert action is given by
\begin{equation}
S_{EH}=\frac{1}{16\pi}\int d^{4}x\sqrt{-g}R,
\end{equation}
$g=\det(g_{\mu\nu})$, and $R$ is the Ricci scalar. The
PP action is given by
\begin{equation}
S_{pp}=\int\sqrt{-g_{\mu\nu}v^{\mu}v^{\nu}}d\tau \left(-\mu+C_{R}R(\left(x(\tau)\right) -C_{V}R_{\mu\nu}\left(x(\tau)\right)\frac{v^{\mu}v^{\nu}}{g_{\mu\nu}v^{\mu}v^{\nu}}\right).
\label{spp}
\end{equation}
The velocity of the point-mass is $v^{\mu}=\frac{dx^{\mu}}{d\tau}$, $\sqrt{-g_{\mu\nu}v^{\mu}v^{\nu}}d\tau$ is the proper time
along the worldline, $R_{\mu\nu}$ is the Ricci tensor, $x(\tau)$
is the worldline position, and $C_{R}$ and $C_{V}$ are the coefficients
of the non-linear terms.  During the early inspiral $v^{2}= v^{\mu}v_{\mu}$ is the small parameter used in the PN expansion. The leading order term describes the motion of the object
along geodesic trajectories of the metric $g_{\mu\nu}$. The two additional terms in Eq.~(\ref{spp})
are proportional to the equations of motion and are therefore redundant.
Thus, we set $C_{R}$ and $C_{V}$  to zero,
\begin{equation}
S_{pp}=-\mu\int d\tau\sqrt{-g_{\mu\nu}v^{\mu}v^{\nu}.}
\end{equation}
With a $\left(-+++\right)$ signature, for timelike worldlines $g_{\mu\nu}dx^{\mu}dx^{\nu}<0$,
the minus sign ensures the square root is real and also that the correct non-relativistic limit is reproduced,
\begin{equation}
S_{pp}\sim\int d\tau\left(-\mu+\frac{1}{2}\mu v^{2}+...\right).
\end{equation}
In our units mass $\sim [L]$ and $\tau\sim[L]$, so $S_{pp}\sim [L^2]$.

The metric terms $H^{BG}_{\mu\nu}$ and $H^{I}_{\mu\nu}$
are not necessarily small on their respective length scales, $r$
and $R_{S}$, but must be small at large distances to ensure an asymptotically
flat metric in the weak-field zone. The metric terms grow as $\sim x^{2}$, which seems to lead to a divergence at large distances. However, this scaling is modified in the weak-field zone upon an appropriate rescaling of the spatial coordinates
$x^{i}$.

\section{Finite-Size terms}

To describe the dynamical interaction between the orbital motion and the internal degrees of freedom of the UCO within an EFT framework, we include finite-size (FS) terms. These are covariant operators that are localized on the worldline of the point-like mass and respect the symmetries of the underlying point-particle action. In the following, we restrict our attention to the electric-type quadrupolar contributions, corresponding to $l=2$. These terms take the form
\begin{equation}
S_{FS}=\int d\tau\ \frac{1}{4}C_2^{E}\left[E_{\mu\nu}E^{\mu\nu}\right],
\label{fs1}
\end{equation}
where $C_2^{E}$ scales as $\left[L^{5}\right]$, which ensures that $S_{FS}\sim [L^2]$. The coefficient $C_2^E$ is the electric quadrupolar (dimensional) Love number (See, for example, \cite{Porto:2016pyg}).

The electric part of the curvature tensor of the gravitational fields is defined as
\begin{equation}
E_{\mu\nu}=R_{\mu\alpha\nu\beta}\left(g\right)v^{\alpha}v^{\beta},
\end{equation}
where the dominant contribution is obtained by taking $\alpha=0,$  $\beta=0$, since $v^0=1,$ while $v^i\sim\sqrt{v^2}$.

When the metric components are allowed to be time-dependent, we need to determine which finite-size terms can be added to the action. In the static case, this is known. In the time-dependent case, we show next that, to the leading PN order, the resulting FS terms are the same as those in the static case.

Due to the antisymmetry of the Riemann tensor, whenever one of the electric moment indices is temporal ($\mu=0$ or $\nu=0$), there is a repetition of indices within an antisymmetric
pair, implying that the corresponding component vanishes.
When both the indices are spatial,
\begin{equation}
R_{i0j0}=\frac{1}{4}\partial_i\left( H^{BG}_{00}+H^{I}_{00}\right)\partial_j \left(H^{BG}_{00}+H^{I}_{00} \right) -\frac{1}{2}\partial_j\partial_j\left(H^{BG}_{00}+H^{I}_{00}. \right),
\end{equation}
Recall that the superscripts $BG$ and $I$ denote background and induced quantities, respectively.

Expanding the electric components of each term in the metric in the weak-field zone to linear order, we find
\begin{align}
& E^{Minkowski}_{ij}  =0,\\
& E^{BG}_{ij}  =R_{i0j0}\left(H^{BG}\right)=-Q_{ij}^{BG},\\
& E^{I}_{ij}  =R_{i0j0}\left(H^{I}\right)=-C_IQ_{ij}^{I}\sim -\frac{1}{r^{5}}Q^{I}_{ij}.
\end{align}

This allows us to write the leading order FS action,
\begin{equation}
S_{FS}=\int d\tau\ \frac{1}{4}C_{E}\left[Q^{BG}_{ij}Q^{BG\ ij}+2\alpha\frac{1}{r^{5}}Q^{BG}_{ij}Q^{I\ ij}+\beta\frac{1}{r^{10}}Q^{I}_{ij}Q^{I\ ij}\right],
\end{equation}
where $\alpha$ and $\beta$ are dimensionless coefficients.

We may redefine $\widetilde{Q}^{I}_{ij}=\alpha Q^{I}_{ij}\to Q^{I}_{ij}$ and.
After absorbing the remaining normalization by the redefinition $\widetilde{\beta}=\beta/\alpha^{2}\rightarrow\beta$ , the FS action become the following,
\begin{equation}
S_{FS}=\int d\tau\ \frac{1}{4}C_{E}\left[Q^{BG}_{ij}Q^{BG\ ij}+2\frac{1}{r^{5}}Q^{BG}_{ij}Q^{I\ ij}+\beta\frac{1}{r^{10}}Q^{I}_{ij}Q^{I\ ij}\right].\label{eq:FS}
\end{equation}
The first term corresponds to a shift in the orbital energy, the second term is an orbit-UCO coupling term, and the third term accounts for the induced energy of the UCO\footnote{The first term could be ignored because it is a background correction, independent of the induced quadrupole. We keep it for completeness.}

In addition to the time-independent terms in the action, we can include terms which depend on time derivatives of $Q^I$. Assuming that the quadrupole moments
associated with the internal structure of the compact object can vary
on their own timescale, which is distinct from that of the orbital quadrupole, there
is no symmetry reason to exclude such terms. In particular,
one can include terms of the form
\begin{equation}
S_{Ik}=\int d\tau\ C_{Ik}\dot{Q}^{I}_{ij}\dot{Q}^{I\ ij},
\end{equation}
where $d/d\tau$ is denoted by a dot.
Since $C_{Ik}$ is the coefficient of the kinetic term of the induced mass quadrupoles it must be positive. Additionally, $\dot{Q}^{I}_{ij}\sim[L^2]$, $S_{Ik}\sim [L^2]$ and $C_{Ik}\sim \left[L^{-3}\right]$.

The sign of the potential term is not discussed
in the literature since it is irrelevant when considering only
the static limit. In Lorentzian spacetime the potential term in
the Lagrangian has a negative sign. Consequently, the
coefficient of the quadrupolar term is negative,
\be
C_{E}<0.
\ee

Focusing on the sector of the action that depends on $Q^{I}_{ij}$,
we write
\begin{equation}
S_{I}=\int d\tau\ \left[C_{Ik}\dot{Q}^{I}_{ij}\dot{Q}^{I\ ij}-\frac{1}{4}\left|C_{E}\right|\beta\frac{1}{r^{10}}Q^{I}_{ij}Q^{I\ ij}-\frac{1}{2}\left|C_{E}\right|\frac{1}{r^{5}}Q^{BG}_{ij}Q^{I\ ij}\right],
\end{equation}
which can be recast as
\bea
 S_{I}=\int d\tau && 2C_{Ik}\Biggl[\biggl(\frac{1}{2}\dot{Q}^{I}_{ij}\dot{Q}^{I\ ij}
-\frac{\omega^{2}_{2}}{2}Q^{I}_{ij}Q^{I\ ij}\biggr)-\frac{\omega^{2}_{2}}{\beta}r^{5}
Q^{BG}_{ij}Q^{I\ ij}\Biggr],
\label{eq:induced action}
\eea
where we have defined,
\begin{equation}
\omega^{2}_{2}=\frac{\beta}{4}\frac{\left|C_{E}\right|}{C_{Ik}}\frac{1}{r^{10}}.
\label{w2defC}
\end{equation}
The characteristic frequency of the internal quadrupolar modes is then denoted by $\omega_2$.

The frequency $\omega_{2}$ has units of $\left[L^{-1}\right]$ and
is expected to be approximately real, as will be discussed in Sect.~\ref{int-frequency}. We
therefore require that the dimensionless coefficient $\beta$ be positive,
\begin{equation}
\beta>0.
\end{equation}

Later, in Sect.~\ref{statk2}, we will show that this implies that the dimensionless electric quadrupolar Love number is negative
\begin{equation}
k_{2}<0
\end{equation}
and discuss the physical implications of this negative sign.

The action $S_{I}$ is seen to be that of a set of linearly coupled harmonic oscillators, capturing the linear response of the internal quadrupole modes to the external tidal fields. This action is
analogous to the harmonic oscillator system discussed in Appendix A, and we therefore
follow similar steps to the ones described there.

In order to obtain the effective action for $Q^{BG}$, we integrate
out the induced quadrupole
\begin{equation}
\left(Q^{I}_{ij}\right)_{ret}(\tau)=\int d\tau'G^{kl}_{ret,\ ij} (\tau-\tau') Q^{BG}_{kl}(\tau'). \label{eq:induced quadrupole}
\end{equation}
The retarded Green function in the frequency domain $G^{kl}_{ret,\ ij}(\Omega)$, can be inferred from the action (\ref{eq:induced action}) by solving the equations of motion,
\begin{equation}
G^{kl}_{ret,\ ij}(\Omega)=\lim_{\epsilon\rightarrow0}\frac{\omega^{2}_{2}r^{5}/\beta} {\left(\Omega-\omega_{2}+i\epsilon\right)\left(\Omega+\omega_{2}+i\epsilon\right)}\delta^{kl}_{ij}. \label{eq:retardedGreenFunction}
\end{equation}
This process will be discussed in detail in Sections \ref{statk2}-\ref{res-k2}.

\section{Conditions on the internal frequency}
\label{int-frequency}

We are considering the inspiral phase of a binary system. This phase ends
when the two companions reach the innermost stable circular orbit (ISCO). It follows that we need to constrain the orbital frequency to remain below the ISCO frequency,
\begin{equation}
\Omega <\Omega_{ISCO}.
\end{equation}
For the non-spinning case, this implies the
bound
\begin{equation}
\Omega<\frac{1}{\sqrt{216}}\frac{1}{M}\sim \frac{1}{7} \frac{1}{R_S},
\end{equation}
while for binaries containing spinning companions this upper bound
relaxes as the ISCO is reached when the distance of the companions is smaller.

Recall that $\omega_{2}$ is the characteristic frequency associated
with the mass quadrupole degree of freedom describing the internal
structure of the compact object $Q^{I}_{ij}$. This frequency is set by the compact
object length scale and therefore scales as $\omega_{2}\propto R^{-1}_{S}$.
However, we will also consider the case in which $\omega_2$ is non-relativistic, as discussed in the introduction,
$
\omega_{2}\ll 1/R_{S}.
$
Recall that this is to be expected for a UCO due to the large  Tolman redshift factor at the surface of the UCO.

In addition, we assume that the imaginary part of  $\omega_{2}$ is parametrically smaller than its real part, namely, that the dissipative component of the internal quadrupolar dynamics can
be neglected. This assumption is justified by the fact that the dominant
dissipative contributions arise from tidal friction effects, which
under reasonable assumptions, can be shown to be negligible compared
to the leading order GW emission, and therefore can be ignored.

If $\omega_2$ is non-relativistic, the internal quadrupole can be resonantly excited by the orbital motion in the inspiral phase. The resonance condition is the following,
\begin{equation}
2\Omega=\omega_{2}.
\end{equation}
This requires that $\omega_2< \dfrac{1}{4} \dfrac{1}{R_S}$, which is not a very restrictive requirement.

\section{Static case}
\label{statk2}

Throughout the early Inspiral phase, the orbital frequency $\Omega$ is small in comparison to other
time scales of the system. Here we begin to formulate an effective action that accounts for time-dependent dynamics, starting with the static case.

In the frequency domain, this corresponds to setting $\Omega=0$.
The retarded Green function is then proportional
to a delta function
\begin{equation}
G^{kl}_{ret,\ ij}(\tau-\tau')=-\frac{r^{5}}{\beta}\delta(\tau-\tau')\delta^{kl}_{ij}.
\end{equation}
In this case Eq.~(\ref{eq:induced quadrupole}) reads
\begin{equation}
Q^{I}_{ij}(\tau)=-\frac{r^{5}}{\beta}Q^{BG}_{ij}(\tau).
\label{InducedQ-eq}
\end{equation}
From the linear response relation\footnote{The superscript $[0]$  indicates the order of the coefficient
in the Taylor expansion of the retarded Green function in the frequency domain about $\Omega=0$.}
\begin{equation}
Q^{I}_{ij}=-\lambda^{[0]}_{2}Q^{BG}_{ij}.
\end{equation}
The dimensionless electric quadrupolar tidal Love number $k_2$ is related to $\lambda^{[0]}_{2}$,
\begin{equation}
\lambda^{[0]}_{2}=-\frac{2}{3}R^{5}_{S}k_{2}.
\end{equation}
We therefore find that
\be
\frac{1}{\beta}=-\frac{2}{3}\left(\frac{R_{S}}{r}\right)^{5}k_{2}.
\label{betak2}
\ee
As expected, $\beta$ is dimensionless and positive and $k_2$ is negative. Furthermore, since $R_{S}/r\sim v^{2}$,
it is of order $5$PN. The precise
order in the PN expansion, however, depends on the complete form of the operator
appearing in the effective action. Substituting the expression for $Q^{I}_{ij}$ from Eq.~(\ref{InducedQ-eq}) and the expression for $\beta$ from Eq.~(\ref{betak2}) into the FS action in  Eq.~(\ref{eq:FS}), we find that
\bea
S_{FS}&=&\int d\tau\ \frac{1}{4}C_{E}\left[Q^{BG}_{ij}Q^{BG\ ij}-\frac{1}{\beta}Q^{BG}_{ij}Q^{BG\ ij}\right]
\cr
&=& \int d\tau\ \frac{1}{4}C_{E} \left[Q^{BG}_{ij}Q^{BG\ ij}+\frac{2}{3}k_{2}\left(\frac{R_{S}}{r}\right)^{5}Q^{BG}_{ij}Q^{BG\ ij}\right].
\label{eq:static_FS_action}
\eea
Here we can see explicitly that the tidal deformation effect for the $\Omega=0$ case
is a correction that enters at the $5$PN order.

Comparing Eq.~(\ref{eq:static_FS_action}), to the standard definition of  the (dimensional) electric quadrupolar Love number $C_2^E$ in the EFT framework in Eq.~(\ref{fs1}),
\be
S_{FS}=\int d\tau\ \left[ \cdots +\frac{1}{4} k_2 R_S^5 Q^{BG}_{ij}Q^{BG\ ij} +\cdots\right],
\ee
As we explained, in our case, the sign of the potential term in the action needs to be negative.
It follows that
\be
-\frac{1}{4} C_E~ \frac{2}{3}k_{2}\left(\frac{R_{S}}{r}\right)^{5}= \frac{1}{4} k_2 R_S^5.
\ee
Thus, we are able to fix $C_E$,
\be
C_E=-\frac{3}{2} r^5.
\label{CEk2}
\ee

To summarize, in the static case, the finite-size action is given by
\bea
S_{FS}
&=&- \int d\tau\ r^5 \left[\frac{3}{8}Q^{BG}_{ij}Q^{BG\ ij}+\frac{1}{4}k_{2}\left(\frac{R_{S}}{r}\right)^{5}Q^{BG}_{ij}Q^{BG\ ij}\right].
\label{eq:static_FS-final}
\eea

\section{Perturbative Time-Dependence Case}
\label{pert-k2}

The next step is to consider the case for which the time dependence is perturbatively small. In the frequency domain, this corresponds to the approximation $\Omega\ll\omega_{2}$ in Eq.~(\ref{eq:retardedGreenFunction}). We Taylor expand the Green function in
the frequency domain around $\Omega=0$,
\begin{equation}
G^{kl}_{ret,\ ij}(\Omega)\cong-\frac{r^{5}}{\beta}\lim_{\epsilon\rightarrow0} \left[1+\frac{i\epsilon\Omega+\Omega^{2}}{\omega^{2}_{2}}+\mathcal{O}(\Omega^{3})\right].
\end{equation}
The expansion may be parametrized as
\begin{equation}
G^{kl}_{ret,\ ij}(\Omega)=-\left(\lambda^{[0]}_{2} +i\lambda^{[1]}_{2}(\Omega)+\lambda^{[2]}_{2}(\Omega^{2})+\mathcal{O}(\Omega)^{3}\right)\delta^{kl}_{ij},
\end{equation}
where $\lambda^{[1]}_{2},\ \text{and}\ \lambda^{[2]}_{2}$ are additional
Wilson coefficients. The minus sign reflects the reduction in the orbital energy due to the coupling to the internal quadrupole. The zeroth order coefficient $\lambda^{[0]}_{2}$ corresponds to the static case. The first order correction,
\begin{equation}
\lambda^{[1]}_{2}=\lim_{\epsilon\rightarrow0}\frac{\epsilon r^{5}}{\beta\omega^{2}_{2}}\Omega,
\end{equation}
corresponds to the dissipative/absorptive contribution. It is imaginary
and odd in $\Omega$, as expected for a retarded response. In the
strict $\epsilon\rightarrow0^{+}$ prescription, this term only encodes
causality, whereas in a physical response function it would be replaced
by a finite dissipative coefficient, e.g. $i\gamma\Omega$.
We assume that the damping rate is
significantly smaller than $\omega_{2}$ and therefore we neglect
it throughout this work. The next real contribution appears at second
order in the expansion, which is also time reversal symmetric and therefore
corresponds to a conservative contribution,
\begin{equation}
\lambda^{[2]}_{2}=\frac{r^{5}}{\beta\omega^{2}_{2}}\Omega^{2}.
\end{equation}
It is the leading time-dependent FS correction beyond
the static linear response. In the time domain, this corresponds to operators
of a higher order and higher powers of time derivatives acting on the tidal field,
\begin{equation}
G^{kl}_{ij}(\tau-\tau')\simeq-\frac{r^{5}}{\beta} \left(1-\frac{1}{\omega^{2}_{2}}\frac{d^{2}}{d\tau^{2}}+\cdots\right)  \int\frac{d\Omega}{2\pi}e^{-i\Omega\left(\tau-\tau'\right)}\delta^{kl}_{ij}.
\end{equation}
The induced quadrupole moment in Eq. (\ref{eq:induced quadrupole})
then becomes
\begin{equation}
\left(Q^{I}_{ij}(\tau)\right)=-\frac{r^{5}}{\beta}Q^{BG}_{ij}(\tau)+ \frac{r^{5}}{\beta\omega^{2}_{2}}\frac{d^{2}}{d\tau^{2}}Q^{BG}_{ij}(\tau) +\cdots.
\end{equation}
Substituting this result into Eq. (\ref{eq:FS}), the FS
action in the perturbative time-dependent case becomes the following,
\bea
S_{FS}&=&\int d\tau\ \frac{1}{4}C_{E}\Biggl[Q^{BG}_{ij}Q^{BG\ ij}
-\frac{1}{\beta}Q^{BG}_{ij}Q^{BG\ ij} \cr
&+& 2\frac{1}{\beta\omega^{2}_{2}}Q^{BG}_{ij}\frac{d^{2}Q^{BG\ ij}}{d\tau^{2}}
-\frac{1}{\beta}\frac{1}{\omega^{4}_{2}}\frac{d^{2}Q^{BG}_{ij}}{d\tau^{2}}
\frac{d^{2}Q^{BG\ ij}}{d\tau^{2}}+\cdots\Biggr].
\eea
Using the expressions for $\beta$ in Eq.~(\ref{betak2}) and for $C_E$ in Eq.~(\ref{CEk2}), and integrating by parts, we may rewrite the FS action as follows,
\bea
S_{FS}&=&-\int d\tau\ r^5 \Bigg\{\frac{3}{8} Q^{BG}_{ij}Q^{BG\ ij}
\cr &+&
\frac{1}{4} k_2 \left(\frac{R_{S}}{r}\right)^{5}~\Bigg[
Q^{BG}_{ij}Q^{BG\ ij}
+ 2\frac{1}{\omega^{2}_{2}} \dot{Q}^{BG}_{ij}\dot{Q}^{BG\ ij}
+\frac{1}{\omega^{4}_{2}} \ddot{Q}^{BG}_{ij}\ddot{Q}^{BG\ ij}+\cdots \Bigg]\Bigg\}.
\eea
The $\cdots$ denote additional
higher time-derivative conservative terms. All the higher-derivative terms enter at PN orders higher than 5PN, since $\dfrac{1}{\omega_{2}}\dfrac{d}{d\tau}\sim v$.
\newline
We also observe that all the higher-derivative operators are proportional to $k_2$.

\section{Resonance Case}
\label{res-k2}

We now turn to calculate the effective action when the orbital frequency
becomes comparable to the characteristic frequency of the internal
structure during the inspiral phase. We focus on the case in which
a resonance excitation is reached when the orbit remains
approximately Keplerian, while still within the domain
of validity of the PN expansion. We solve for the retarded Green function in Eq.~(\ref{eq:retardedGreenFunction})
without making a low-frequency approximation. Evaluating the integral
yields
\begin{equation}
G^{kl}_{ret,\ ij}(\tau-\tau')=-\frac{r^{5}}{\beta}\omega_{2}~\sin\left[\omega_{2}\left(\tau-\tau'\right)\right]\theta\left(\tau-\tau'\right) \delta^{kl}_{ij},
\end{equation}
and the induced quadrupole moment in Eq. (\ref{eq:induced quadrupole})
becomes
\begin{equation}
Q^{I}_{ij}(\tau)=-\frac{r^{5}}{\beta}\int d\tau'\ \omega_{2}~\sin\left[\omega_{2}\left(\tau-\tau'\right)\right]\theta\left(\tau-\tau'\right)Q^{BG}_{ij}(\tau'). \label{eq:RetIndQP}
\end{equation}
To calculate the contribution of the potential terms, we substitute
the result in Eq.~(\ref{eq:RetIndQP}) into the action (\ref{eq:FS}), taking into account the proper-time dependence of the tensors explicitly,
\bea
&& \ \hspace{-.5in} S_{FS}=\int d\tau\ \frac{1}{4}C_{E}\Bigg[Q^{BG}_{ij}(\tau)Q^{BG\ ij}(\tau)
\label{eq:ResAction}  \\
&& \ \hspace{-.5in} -\frac{2}{\beta}Q^{BG}_{ij}(\tau)\int d\tau'\ \omega_{2}\sin\left[\omega_{2}\left(\tau-\tau'\right)\right]
\theta(\tau-\tau') Q^{BG\ ij}(\tau')
\cr
&& \ \hspace{-.5in} +\frac{1}{\beta}\int d\tau'\ \omega_{2}~\sin\left[\omega_{2}\left(\tau-\tau'\right)\right]\theta(\tau-\tau')Q^{BG}_{ij}(\tau')\int d\tau''\ \omega_{2}\sin\left[\omega_{2}\left(\tau-\tau''\right)\right]\theta(\tau-\tau'')Q^{BG\ ij}(\tau'')\Bigg]. \nonumber
\eea
The second term in Eq.~(\ref{eq:ResAction}) comes from the coupling term in the action and the third term comes from the potential term.

In the coupling term, we first perform the orbital background contraction described in Appendix B,
\bea
L_{coupling}(\tau)=\left(-\frac{3}{\beta}\right)\int d\tau' && \Biggl[\omega_{2}\sin\left[\omega_{2}\left(\tau-\tau'\right)\right]\theta(\tau-\tau') \times \cr && \left(\cos^{2}\left[\Omega(\tau-\tau')\right]-\frac{1}{3}\right)\Biggr]
Q^{BG}_{ij}(\tau)Q^{BG\ ij}(\tau).
\eea
Then, using integration by parts, as described in Appendix C,
\bea
&& L_{coupling}(\tau)= \left(-\frac{3}{\beta}\right)\int d\tau'\Bigg[\delta(\tau-\tau')\left(\cos^{2}\left[\Omega(\tau-\tau')\right]-\frac{1}{3}\right)
\cr
&-&\theta(\tau-\tau')\Omega\sin\left[2\Omega\left(\tau-\tau')\right)\right]\Bigg] \cos\left[\omega_{2}\left(\tau-\tau'\right)\right]Q^{BG}_{ij}(\tau)Q^{BG\ ij}(\tau). \ \ \
\eea
The integration over the delta function  contributes to reproducing the static case result, even when the
induced quadrupole in Eq.(\ref{eq:RetIndQP}) is constructed
using a contour integral which enforces causality $\tau-\tau'>0$.
Using the identity $\cos a\sin b=\frac{1}{2}\left(\sin(a+b)+\sin(b-a)\right)$,
we obtain
\bea
&&L_{coupling}(\tau)=-\frac{2}{\beta} Q^{BG}_{ij}(\tau)Q^{BG\ ij}(\tau)
\\
&+&\frac{3}{2\beta}\int^{\tau}_{-\infty}d\tau'\ \Omega\Biggl[\sin\left[\left(2\Omega+\omega_{2}\right)(\tau-\tau')\right]
+\sin\left[\left(2\Omega -\omega_{2}\right)(\tau-\tau')\right]\Biggr]
Q^{BG}_{ij}(\tau)Q^{BG\ ij}(\tau). \nonumber
\eea

We impose the resonance condition
\begin{equation}
2\Omega(\tau'=\tau_{R})=\omega_{2},
\label{eq:ResCondition}
\end{equation}
which defines the time of resonance $\tau_{R}$ and, as detailed in Appendix D, we use the stationary phase approximation to evaluate the integral. The result is the following,
\bea
&&L_{coupling}(\tau)=-\frac{2}{\beta}Q^{BG}_{ij}(\tau)Q^{BG\ ij}(\tau)
\\ &+& \frac{3\sqrt{\pi}}{4\sqrt{2}\beta}\theta\left(\tau-\tau_{R}\right)\frac{\omega_{2}}{\sqrt{\dot{\Omega}_{R}}} \left(1+\frac{\dot{\Omega}_{R}}{\omega_{2}}\left(\tau-\tau_{R}\right)\right) \sin\left[\frac{\dot{\Omega}_{R}}{2}\left(\tau-\tau_{R}\right)^2-\frac{\pi}{4}\right] Q^{BG}_{ij}(\tau)Q^{BG\ ij}(\tau), \nonumber
\eea
where the rate of change of the orbital frequency at the time of resonance is denoted by $\dot{\Omega}_{R}=\dot{\Omega}\left(\tau=\tau_{R} \right)$.
Note that $\Omega = \Omega (\tau)$ varies slowly with time, so the term proportional to $\dot{\Omega}$ can be neglected (See Appendix~\ref{subsec:(E)-Omega-Dot} for details).

To evaluate the potential term of the induced quadrupole, we first consider
the contraction of the background quadrupoles according to Appendix B,
\begin{equation}
Q^{BG}_{ij}(\tau')Q^{BG\ ij}(\tau'')=\frac{3}{2}\left(\cos^{2}\left[\Omega(\tau'-\tau'')\right] -\frac{1}{3}\right)Q^{BG}_{ij}(\tau)Q^{BG\ ij}(\tau).
\end{equation}
The trigonometric relation
$
\cos^{2}\left(a-b\right)=\frac{1}{2}+\frac{1}{2}\cos2a\cos2b+\frac{1}{2}\sin2a\sin2b
$,
leads to
\bea
Q^{BG}_{ij}(\tau')Q^{BG\ ij}(\tau'') =\frac{3}{4}\Bigg(\frac{1}{3}+\cos\left[2\Omega\left(\tau-\tau'\right)\right] \cos\left[2\Omega\left(\tau-\tau''\right)\right]+
\cr
+\sin\left[2\Omega\left(\tau-\tau'\right)\right]\sin\left[2\Omega\left(\tau-\tau''\right)\right]\Bigg)Q^{BG}_{ij}(\tau)Q^{BG\ ij}(\tau).
\eea
Now we can rewrite the contribution from the potential term in the following way,
\begin{equation}
L_{potential}(\tau)=\frac{3}{4\beta}\left[I^{2}_{1}+I^{2}_{2}+I^{2}_{3}\right]Q^{BG}_{ij}(\tau)Q^{BG\ ij}(\tau),\label{eq:InducedPotential}
\end{equation}
defining the three integrals,
\begin{equation}
I_{1}=\int d\tau'\ \omega_{2}\sin\left[\omega_{2}\left(\tau-\tau'\right)\right] \theta(\tau-\tau')\cos\left[2\Omega\left(\tau-\tau'\right)\right],\label{eq:I1}
\end{equation}
\begin{equation}
I_{2}=\int d\tau'\ \omega_{2}\sin\left[\omega_{2}\left(\tau-\tau'\right)\right] \theta(\tau-\tau')\sin\left[2\Omega\left(\tau-\tau'\right)\right]
\end{equation}
and
\begin{equation}
I_{3}=\int d\tau'\ \frac{1}{\sqrt{3}}\omega_{2}\sin\left[\omega_{2}\left(\tau-\tau'\right)\right]\theta(\tau-\tau').
\end{equation}
Integrating by parts, as described in Appendix C, the integrals $I_1$, $I_2$ and $I_3$
can be evaluated,
\begin{equation}
I_{1}=\int d\tau'\left[\delta\left(\tau-\tau'\right)\cos\left[2\Omega\left(\tau-\tau'\right)\right] -2\Omega\theta(\tau-\tau')\sin\left[2\Omega\left(\tau-\tau'\right)\right]\right] \cos\left[\omega_{2}\left(\tau-\tau'\right)\right],
\end{equation}
\begin{equation}
I_{2}=\int d\tau'\left[\delta\left(\tau-\tau'\right)\sin\left[2\Omega\left(\tau-\tau'\right)\right] +2\Omega\theta(\tau-\tau')\cos\left[2\Omega\left(\tau-\tau'\right)\right]\right] \cos\left[\omega_{2}\left(\tau-\tau'\right)\right],
\end{equation}
and
\begin{equation}
I_{3}=\int d\tau'\ \frac{1}{\sqrt{3}}\delta\left(\tau-\tau'\right)\cos\left[\omega_{2}\left(\tau-\tau'\right)\right].
\end{equation}
The terms containing a factor of a delta function in $I_{1}$ and $I_{3}$ contribute to reproducing
the static case result. For $I_{1}$ and $I_{2}$ we use the trigonometric identities
$
2\sin a\cos b=\sin\left(a+b\right)+\sin\left(a-b\right)
$
and
$
2\cos a\cos b=\cos\left(a+b\right)+\cos\left(a-b\right).
$
This will allow us to use the resonance condition (\ref{eq:ResCondition})
and the stationary phase approximation, as described in appendix D.
The resulting integrals are found to be
\begin{equation}
I_{1}=1-\frac{\sqrt{\pi}}{2\sqrt{2}\beta}\theta\left(\tau-\tau_{R}\right) \frac{\omega_{2}}{\sqrt{\dot{\Omega}_{R}}}\left(1+\frac{\dot{\Omega}_{R}}{\omega_{2}} \left(\tau-\tau_{R}\right)\right)\sin\left[\frac{\dot{\Omega}_{R}}{2}\left(\tau-\tau_{R}\right)^2 -\frac{\pi}{4}\right],
\end{equation}
\begin{equation}
I_{2}=\frac{\sqrt{\pi}}{2\sqrt{2}\beta}\theta\left(\tau-\tau_{R}\right) \frac{\omega_{2}}{\sqrt{\dot{\Omega}_{R}}}\left(1+\frac{\dot{\Omega}_{R}}{\omega_{2}} \left(\tau-\tau_{R}\right)\right)\cos\left[\frac{\dot{\Omega}_{R}}{2}\left(\tau-\tau_{R}\right)^2 -\frac{\pi}{4}\right]
\end{equation}
and
\begin{equation}
I_{3}=\frac{1}{\sqrt{3}}.
\end{equation}
Substituting the results of the integrals into Eq.~(\ref{eq:InducedPotential}),  the final result for the potential term  is given by
\bea
&& L_{potential}(\tau)=\frac{1}{\beta}Q^{BG}_{ij}(\tau)Q^{BG\ ij}(\tau)
\cr
&-&\frac{3\sqrt{\pi}}{4\sqrt{2}\beta} \theta\left(\tau-\tau_{R}\right) \frac{\omega_{2}}{\sqrt{\dot{\Omega}_{R}}} \left(1+\frac{\dot{\Omega}_{R}}{\omega_{2}}\left(\tau-\tau_{R}\right)\right) \sin\left[\frac{\dot{\Omega}_{R}}{2}\left(\tau-\tau_{R}\right) -\frac{\pi}{4}\right]Q^{BG}_{ij}(\tau)Q^{BG\ ij}(\tau)
\cr &+&
\frac{3\pi}{32\beta}\frac{\omega^{2}_{2}}{\dot{\Omega}_{R}} \left(1+\frac{\dot{\Omega}_{R}}{\omega_{2}}\left(\tau-\tau_{R}\right)\right)^{2} \theta\left(\tau-\tau_{R}\right)\Bigg]Q^{BG}_{ij}(\tau)Q^{BG\ ij}(\tau).
\eea
Here, the first term contributes to the static result, the second term cancels with the time-dependent contribution from the coupling term, and the third term contributes to the FS term correction in the time-dependent EFT action.

We now proceed to express the action in terms of the Schwarzschild radius $R_S$ and the binary mass ratio $q$. First, we rewrite Kepler's third law,
\begin{equation}
\Omega^{2}=\frac{1}{2}(1+q)\left(\frac{R_{S}}{r}\right)\frac{1}{r^{2}}.
\label{om2}
\end{equation}
Then, express the quadrupole formula for radial decay for quasi-circular binaries,
\begin{equation}
\dot{r}=-\frac{8}{5}q\left(1+q\right)\left(\frac{R_{S}}{r}\right)^{3}.
\end{equation}
Taking the time derivative of Eq.~(\ref{om2}),
\begin{equation}
\dot{\Omega}=-\frac{3}{2}\Omega\frac{\dot{r}}{r},
\end{equation}
we can express the ratio $\frac{\dot{\Omega}}{\Omega^{2}}$ in terms of $R_S$ and $q$,
\begin{equation}
\frac{\dot{\Omega}}{\Omega^{2}}=\frac{12\sqrt{2}}{5}q\sqrt{1+q}\left(\frac{R_{S}}{r}\right)^{5/2}.
\end{equation}
This ratio is of order 2.5PN, which means that the resonance correction
first appears at the 2.5PN order, lower than the static correction, which is of 5PN order.

The term $\frac{\dot{\Omega}_{R}}{\omega_{2}}\left(\tau-\tau_{R}\right)$ is subleading with respect to the leading term $1$ and we therefore neglect it in what follows.

Summarizing our results this far, we find that the FS effective action (\ref{eq:FS}) is given by,
\begin{equation}
S_{FS}=\int d\tau\ \frac{1}{4}C_{E}\left[Q^{BG}_{ij}Q^{BG\ ij}-\frac{1}{\beta}Q^{BG}_{ij}Q^{BG\ ij}+\frac{3\pi}{32\beta}\frac{\omega^{2}_{2}}{\dot{\Omega}_{R}} \theta\left(\tau-\tau_{R}\right)Q^{BG}_{ij}Q^{BG\ ij}\right].
\end{equation}

To obtain the final contribution to the FS action from the potential and coupling terms, for $q \ll 1$, we approximate the numerical prefactor, substitute $\beta$  from Eq.~(\ref{betak2}) and $C_E$ from Eq.~(\ref{CEk2}),
\bea
S_{FS}\simeq - \int &d\tau& r^5 \Biggl[\frac{3}{8}+\frac{1}{4}\left(\frac{R_{S}}{r}\right)^{5}k_{2}
\cr && -
\left(\frac{9}{400}\frac{1}{q}\left(\frac{r_R}{r}\right)^{5/2} \left(\frac{R_{S}}{r}\right)^{5/2}k_{2}\right) \theta\left(\tau-\tau_{R}\right)\Biggr]Q^{BG}_{ij}Q^{BG\ ij}.
\label{eq:dynamic_FS_action}
\eea
Here, we defined the orbital separation at time of resonance $r_R= r(\tau=\tau_{R})$.
For any $\tau>\tau_{R}$ in the inspiral phase,  $\frac{r_R}{r}>1$.
For future reference, we write the leading term in the action,
\be
S_{FS,Res}= \frac{9}{400} \frac{1}{q}~ k_{2}   \int d\tau~ r^5~
\left(\frac{r_R}{r}\right)^{5/2} \left(\frac{R_{S}}{r}\right)^{5/2} \theta\left(\tau-\tau_{R}\right)~Q^{BG}_{ij}Q^{BG\ ij}.
\label{fsRes}
\ee

To this result, we need to add the contribution arising from the kinetic energy
of the induced quadrupole.


Writing explicitly the kinetic part, using Eq.~(\ref{eq:RetIndQP}) to integrate out $Q^{I}$,
and Eq.~(\ref{w2defC}) to relate $C_{Ik}$ and $C_E$,
\begin{equation}
C_{Ik}=\frac{1}{4}\left|C_{E}\right|\frac{\beta}{\omega^{2}_{2}~ r^{10}},
\end{equation}
\bea
&&S_{FSk}=\int d\tau\ C_{Ik}\dot{Q}^{I}_{ij}\dot{Q}^{I\ ij}=
\cr
&&=\int d\tau\ \frac{1}{4}|C_{E}|\ \frac{1}{\beta}\left[\frac{1}{\omega_{2}}\frac{d}{d\tau}\int d\tau'\ \omega_{2}\sin\left[\omega_{2}\left(\tau-\tau'\right)\right] \theta(\tau-\tau')Q^{BG}_{ij}(\tau')\right]
\cr
&&\times\left[\frac{1}{\omega_{2}}\frac{d}{d\tau}\int d\tau''\ \omega_{2}\sin\left[\omega_{2}\left(\tau-\tau''\right)\right]\theta(\tau-\tau'')Q^{BG\ ij}(\tau'')\right],
\label{kact}
\eea

The $\tau$ derivatives in Eq.~(\ref{kact}) is given by,
\bea
\frac{\partial}{\partial\tau}\left(\sin\left[\omega_{2}\left(\tau-\tau'\right)\right] \theta\left(\tau-\tau'\right)\right)=
\cr
=\omega_{2}\cos\left[\omega_{2}\left(\tau-\tau'\right)\right] \theta\left(\tau-\tau'\right)+\delta\left(\tau-\tau'\right) \sin\left[\omega_{2}\left(\tau-\tau'\right)\right].
\eea
Here, the second term vanishes after integrating over $\tau'$. Using the results of Appendix B,
the kinetic action becomes
\begin{equation}
S_{FSk} =\int d\tau\ \frac{3}{16}\frac{|C_{E}|}{\beta}\left(I^{2}_{4}+I^{2}_{5} +I^{2}_{6}\right)Q^{BG}_{ij}Q^{BG\ ij},
\label{sfsk}
\end{equation}
where we defined
\begin{equation}
I_{4}(\tau) = \int d\tau'\ \omega_{2}\cos\left[\omega_{2}\left(\tau-\tau'\right)\right] \theta\left(\tau-\tau'\right)\cos\left[2\Omega\left(\tau-\tau'\right)\right],
\end{equation}
\begin{equation}
I_{5}(\tau) =\int d\tau'\ \omega_{2}\cos\left[\omega_{2}\left(\tau-\tau'\right)\right] \theta\left(\tau-\tau'\right)\sin\left[2\Omega\left(\tau-\tau'\right)\right],
\end{equation}
and
\begin{equation}
I_{6}(\tau) =\int d\tau'\ \omega_{2}\cos\left[\omega_{2}\left(\tau-\tau'\right)\right]\theta\left(\tau-\tau'\right)\frac{1}{\sqrt{3}}.
\end{equation}
After integration by parts, as in Appendix C, we find that
\begin{equation}
I_{4}(\tau) =\int d\tau'\ \left(+2\Omega\right)\theta\left(\tau-\tau'\right) \sin\left[2\Omega\left(\tau-\tau'\right)\right]\sin\left[\omega_{2}\left(\tau-\tau'\right)\right],
\end{equation}
\begin{equation}
I_{5}=\int d\tau'\ \left(-2\Omega\right)\theta\left(\tau-\tau'\right) \cos\left[2\Omega\left(\tau-\tau'\right)\right]\sin\left[\omega_{2}\left(\tau-\tau'\right)\right],
\end{equation}
and
\begin{equation}
I_{6}=0.
\end{equation}
Using the trigonometric identities $2\sin a\sin b=\cos\left(a-b\right)-\cos\left(a+b\right)$
and $2\sin a\cos b=\sin\left(a+b\right)+\sin\left(a-b\right)$,
the resonance condition and the stationary phase approximation, we find that $I_4$ is given by
\begin{equation}
I_{4}=\frac{\sqrt{\pi}}{2\sqrt{2}\beta} \theta\left(\tau-\tau_{R}\right)\frac{\omega_{2}}{\sqrt{\dot{\Omega}_{R}}} \left(1+\frac{\dot{\Omega}_{R}}{\omega_{2}}\left(\tau-\tau_{R}\right)\right) \cos\left[\frac{\dot{\Omega}_{R}}{2}\left(\tau-\tau_{R}\right)^2-\frac{\pi}{4}\right],
\label{i4}
\end{equation}
and $I_5$ by
\begin{equation}
I_{5}=\frac{\sqrt{\pi}}{2\sqrt{2}\beta}\theta\left(\tau-\tau_{R}\right) \frac{\omega_{2}}{\sqrt{\dot{\Omega}_{R}}}\left(1+\frac{\dot{\Omega}_{R}}{\omega_{2}} \left(\tau-\tau_{R}\right)\right)\sin\left[\frac{\dot{\Omega}_{R}}{2} \left(\tau-\tau_{R}\right)^2-\frac{\pi}{4}\right].
\label{i5}
\end{equation}

Substituting Eqs.~(\ref{i4}) and (\ref{i5}) into Eq.~(\ref{sfsk}), we obtain the final form of the kinetic part of the action,
\bea
&&S_{FSk}=\int d\tau\ \frac{1}{4}|C_{E}|\left[\frac{3\pi}{32\beta} \frac{\omega^{2}_{2}}{\dot{\Omega}_{R}} \left(1+\frac{\dot{\Omega}_{R}}{\omega_{2}}\left(\tau-\tau_{R}\right)\right)^{2} \theta\left(\tau-\tau_{R}\right)\right]Q^{BG}_{ij}(\tau)Q^{BG\ ij}(\tau)
\cr
&\simeq& -\int d\tau\ \frac{1}{4}|C_{E}|\left[\left(\frac{3}{50q}\left(\frac{r_{R}}{r}\right)^{5/2} \left(\frac{R_{S}}{r}\right)^{5/2}k_{2}\right) \theta\left(\tau-\tau_{R}\right)\right]
Q^{BG}_{ij}(\tau)Q^{BG\ ij}(\tau).
\eea
Substituting the value of $C_E$ from Eq.~(\ref{CEk2}), we arrive at our final result for $S_{FSk}$,
\bea
&&S_{FSk}= -\frac{9}{400}~\frac{1}{q}~k_{2}  \int d\tau\ r^5  \left(\frac{r_{R}}{r}\right)^{5/2} \left(\frac{R_{S}}{r}\right)^{5/2} \theta\left(\tau-\tau_{R}\right)
Q^{BG}_{ij}(\tau)Q^{BG\ ij}(\tau).
\label{fskF}
\eea

Comparing the kinetic part to the potential part (\ref{fsRes}), we see that they have equal magnitude but opposite signs.  This was expected because, on-shell, the magnitudes of the potential and kinetic terms for a simple harmonic oscillator are equal and have opposite signs.  The kinetic term is negative because $k_2$ is negative and the potential term is positive.  The correction to the on-shell action then formally vanishes, as expected, for an on-shell action of a harmonic oscillator.   We are interested, though, in the energy transfer from the orbit to the internal quadrupole, which does not vanish, rather, it is the sum of the kinetic and potential energy corrections, leading to an overall reduction of the orbital energy, as expected. This will be discussed in the next Section.

\section{Energy transfer and phase shift}

\subsection{Energy transfer}

The energy correction to the orbital energy due to resonant excitation of the internal quadrupole is the sum of the kinetic part and the potential part, which is simply equal to twice the kinetic part
in Eq.~(\ref{fskF}),
\begin{equation}
\Delta E_{res}(\tau)=   r^5 ~ \frac{9}{200}\frac{1}{q} k_{2} \left(\frac{r_{R}}{r}\right)^{5/2} \left(\frac{R_{S}}{r}\right)^{5/2} \theta\left(\tau-\tau_{R}\right)
Q^{BG}_{ij}(\tau)Q^{BG\ ij}(\tau),
\end{equation}
with $r(\tau)$ approximated by the Kepler value.
Substituting the Kepler value for $Q^{BG}_{ij}(\tau)Q^{BG\ ij}(\tau)$ from Eq.~(\ref{QBG2}) in Appendix B, we find
\begin{equation}
\Delta E_{res}(\tau)=  - ~ \frac{27}{400}~q~k_{2} \left(\frac{r_{R}}{r}\right)^{5/2} \left(\frac{R_{S}}{r}\right)^{7/2} R_S ~\theta\left(\tau-\tau_{R}\right).
\label{DEres}
\end{equation}
Since $k_2$ is negative, the orbital energy is reduced, as expected.

This energy transfer seems extremely small: it is proportional to a small coefficient and to $q$ and $k_2$, which are both small parameters. However, its significance can only be appreciated when comparing this energy to the radiated energy by GW, which is also small, and when determining how this small energy transfer affects the phase of the radiated energy. This is what we do next.

\subsection{Phase shift}

We evaluate the relative phase shift of the emitted GW induced by the energy transfer following \cite{Goldberger:2004jt},
\be
\Delta\Phi=\frac{4}{R_S}\int d v ~v^3~ \frac{d~ \Delta E_{res}/dv}{P(v)}.
\label{DPhiv}
\ee
Here $P(v)$ is the emitted GW power, which according to the quadrupole formula is given by (recall that we are using units in which $G=1$ and $c=1$),
\be
P(v)=\frac{32}{5} q^2 v^{10}.
\ee
Additionally, using the Kepler relation $\dfrac{R_S}{r}=2 v^2$, the transferred energy in Eq.~(\ref{DEres}) can be expressed as a function of $v$,
\be
\Delta E_{res}(v)=\frac{108}{25}~ q~ k_2 \left(\frac{r_R}{R_S}\right)^{5/2}~ v^{12}~ R_S.
\ee
Using these expressions, we can evaluate the resulting phase shift by performing the integral in Eq.~(\ref{DPhiv}). Calculating the integral from the resonance to the ISCO and taking into account that most of the support of the integral comes from the small $r$, large $v$ region, we may approximate the integral by its value at $v_{ISCO}=\sqrt{\frac{1}{2}\frac{R_S}{r_{ISCO}}}$.   The result is the following,
\be
\Delta\Phi\simeq 1.14~ \frac{1}{q}~ k_2~ \left(\frac{r_R}{r_{ISCO}}\right)^{5/2}.
\label{DPhif}
\ee
Recall that $k_2$ is negative, and that after the resonance the orbital angular frequency is higher than in the Keplerian case, so the Keplerian phase lags the actual phase by the amount in Eq.~(\ref{DPhif}).

Our resulting phase shift has the same parametric dependence on $q$ and $k_2$ as the one found in \cite{Avitan:2023txy}, but the numerical coefficient that we find here is much larger than the one calculated in \cite{Avitan:2023txy}, in part due to the approximations that we used there\footnote{Additionally, there is a factor of 6 missing in \cite{Avitan:2023txy} due to an incorrect evaluation of a normalization factor.}.

\section{Conclusions and Outlook}

We used the point-particle effective field theory framework to discuss effects associated with exciting the internal electric quadrupole of UCOs. We incorporated the time dependence of the internal quadrupole to leading order by adding its quadratic kinetic term in the EFT.  We showed that integrating out the internal quadrupole leads to a series of time-dependent finite-size terms with specific numerical coefficients. We found that the electric quadrupole Love number $k_2$ is negative and that all finite-size terms are proportional to $k_2$. We then discussed resonant excitation of the internal quadrupole by a circular orbital motion and showed that it induces a leading 2.5 Post-Newtonian term, to be compared with the 5 Post-Newtonian term in the time-independent case. We calculated the resonant energy transfer from the orbit to the UCO and found that it is proportional to $k_2$ and inversely proportional to the mass ratio $q$. Finally, we used this result to calculate the phase shift of the emitted GW relative to the Keplerian value.  Our results confirm the expectation that observing resonant excitation during a binary inspiral, particularly for small $q$, offers one of the best ways to determine whether the tidal response of astrophysical black holes vanishes and, if it does not, learn about their internal excitation spectrum.

Our results can be used to calculate the differences between GR predictions for the emitted GW waveform during a binary inspiral and the waveform if one or both binaries are UCOs. Because the effect enters at 2.5PN order, the energy transfer from the orbit to the internal quadrupole and consequently the induced phase shift of the emitted GW is significant. Hopefully, this will lead to the discovery of the internal structure of astrophysical black holes or rule out their existence.

The effective action that we constructed does not rely on a specific UCO model. Consequently, it can describe any model of a horizonless compact object that possesses a spectrum of non-relativistic internal modes. In specific models, we expect $k_2$ and the resonant frequency, which determines the resonance radial position, to be related rather than being independent parameters. This would simplify the relationship between $k_2$ and the phase shift.
A measurement of the phase shift predicted by our analysis could therefore not only indicate a departure from GR predictions but also provide evidence in favor of a specific class of horizonless models sharing these characteristics. Moreover, such an observation could distinguish between models that predict negative tidal Love numbers and those predicting positive ones.

\section*{Acknowledgements}

We thank Misha Smolkin for helping us navigate through the many intricacies of the effective field theory in many discussions.
The research is supported by the German Research Foundation through a German-Israeli Project Cooperation (DIP) grant ``Holography and the Swampland'' and by VATAT (Israel planning and budgeting committee) grant for supporting theoretical high energy physics.  RB thanks the theory department at CERN, the department of theoretical physics at the University of Geneva and the Arnold Sommerfeld Centre at LMU, Munich, for their hospitality.\\

\appendix

{\Large{\bf Appendices}}

\renewcommand{\thesection}{\Alph{section}}

\section{{Finite Size coefficients}
\label{subsec:(A)-coupled HO analoug}}

In this Appendix we show that all the coefficients of the FS terms in the action are related, once the response function of the internal degrees of freedom is known. To this end, we use a simple
example of two coupled harmonic oscillators,  $\phi_{1}$ and $\phi_{2}$ with characteristic frequencies $\omega_{1}$ and $\omega_{2}$, respectively. To obtain the response function, we need only consider the time dependence and ignore any spatial dependence. Additionally, we neglect the effects of damping.
The external gravitational potential will be represented by $\phi_{1}$ and $\phi_{2}$ will represent
the most dominant mode of the internal structure. A coupling term between $\phi_{1}$ and $\phi_{2}$ is included.

The action of the coupled harmonic oscillator is then
given by
\be
S[\phi_{1},\phi_{2}]
=\int~dt~\left[\frac{1}{2}\dot{\phi}^{2}_{1}(t) +\frac{1}{2}\dot{\phi}^{2}_{2}(t)-\frac{1}{2}\omega^{2}_{1}\phi^{2}_{1}(t) -\frac{1}{2}\omega^{2}_{2}\phi^{2}_{2}(t)-C\phi_{1}(t)\phi_{2}(t)\right].
\label{eq:action}
\ee
Our goal is to obtain an effective action for the field $\phi_{1}$. To do so,
we need to integrate out $\phi_{2}$, which we can do exactly. In
general,
\begin{equation}
\phi_{2}(t)=\int F(t-t')\phi_{1}(t')dt'.
\end{equation}

In the frequency domain
\begin{equation}
\phi_{i}(t)=\int\frac{d\omega}{2\pi}\phi_{i}(\omega)e^{i\omega t},
\end{equation}
so the action can be rewritten as
\bea
S&=&\int d\omega\ \frac{1}{4\pi}\Biggl[(\omega^{2}-\omega^{2}_{1})\phi_{1}(\omega)\phi_{1}(-\omega)
\cr &&
+(\omega^{2}-\omega^{2}_{2})\phi_{2}(\omega)\phi_{2}(-\omega)
-C\phi_{1}(\omega)\phi_{2}(-\omega)-C\phi_{2}(\omega)\phi_{1}(-\omega)\Biggr],
\eea
The equation of motion results in an expression for $\phi_{2}(\omega)$ in terms of $\phi_{1}(\omega)$,
\begin{equation}
\phi_{2}(\omega)=\frac{C}{\omega^{2}-\omega^{2}_{2}}\phi_{1}(\omega).
\end{equation}
Using the inverse Fourier transform, one gets an expression for $\phi_{2}(t)$ in terms of $\phi_{1}(t)$,
\begin{equation}
\phi_{2}(\omega)=\frac{C}{\omega^{2}-\omega^{2}_{2}}\int\phi_{1}(t')e^{-i\omega t'}dt',
\end{equation}
\begin{equation}
\phi_{2}(t)=\int dt'\left(\int\frac{d\omega}{2\pi}\frac{C}{\omega^{2}-\omega^{2}_{2}} e^{i\omega(t-t')}\right) \phi_{1}(t').
\label{eq: IntOut}
\end{equation}

To compute the response function, we slightly shift the poles below
the real $\omega$ axis in the complex plane, moving them to $\pm\omega_{2}-i\epsilon$.
This prescription allows us, in a later calculation, to close the
integration contour in the lower half-plane and thereby enforce causality,
$t-t'>0$. With this prescription, the response function is given
by the following integral,
\begin{equation}
F(t-t')=\lim_{\epsilon\rightarrow0}\int\frac{d\omega}{2\pi}\ \frac{C}{\left(\omega-\omega_{2}+i\epsilon\right) \left(\omega+\omega_{2}+i\epsilon\right)}e^{i\omega(t-t')}.
\label{eq:response}
\end{equation}

The static case corresponds to setting
$\omega=0$,
\begin{equation}
\lim_{\epsilon\rightarrow 0}\frac{C}{\omega^{2}-\omega^{2}_{2}+i\epsilon\omega}\Bigg|_{\omega=0} =-\frac{C}{\omega^{2}_{2}}
\end{equation}
and the leading order contribution to Eq.~(\ref{eq: IntOut}) is then
\be
\phi^{[0]}_{2}(t)=-\frac{C}{\omega^{2}_{2}}\int dt'\left(\int\frac{d\omega}{2\pi}e^{i\omega(t'-t)}\right)\phi_{1}(t')=-\frac{C}{\omega^{2}_{2}}\int dt'\ \delta(t-t')\ \phi_{1}(t'),
=-\frac{C}{\omega^{2}_{2}}\phi_{1}(t).
\ee
Now, using an expansion around $\omega=0$ up to second order
in $\omega$, we obtain the subleading contribution
\begin{equation}
\lim_{\epsilon\rightarrow0}\frac{C}{\omega^{2}-\omega^{2}_{2}+i\epsilon\omega}\cong -\frac{C}{\omega^{2}_{2}}\left(1+\frac{\omega^{2}}{\omega^{2}_{2}}\right),
\end{equation}
Substituting this back into Eq.~(\ref{eq: IntOut}), we find
\bea
\phi^{[2]}_{2}(t)&=&-\frac{C}{\omega^{4}_{2}}\int dt'\left(\int\frac{d\omega}{2\pi}\omega^{2}e^{i\omega(t'-t)}\right)\phi_{1}(t')=\frac{C}{\omega^{4}_{2}}\frac{d^{2}}{dt^{2}}\int dt'\left(\int\frac{d\omega}{2\pi}e^{i\omega(t'-t)}\right)\phi_{1}(t')
\cr
&=&\frac{C}{\omega^{4}_{2}}\frac{d^{2}}{dt^{2}}\int dt'\ \delta(t-t')\ \phi_{1}(t')=\frac{C}{\omega^{4}_{2}}\frac{d^{2}}{dt^{2}}\phi_{1}(t).
\eea

However, if we consider the possibility
that, at some point during the evolution of the system, the condition
$\omega\ll\omega_{2}$ is no longer satisfied then the approximation
methods discussed above are no longer applicable. Instead, we must
explicitly evaluate the integral defining the response function. To
solve Eq.~(\ref{eq:response}) exactly, we consider a closed contour
consisting of two parts: the first runs along the real axis, $\omega=x+i0$,
with $x:-R\rightarrow+R$, the second is the semicircle in the lower
half-plane, parameterized by $\omega=R\cos\theta-i\sin\theta$, with
$\theta:0\rightarrow\pi$. In the limit $R\rightarrow\infty$ , the
contribution from the semicircular arc vanishes for $t-t'>0$, so
that the contour integral reduces to the integral along the real axis.

Using the residue theorem the response function in Eq. (\ref{eq:response})
becomes
\begin{equation}
F(t-t')=\lim_{\epsilon\rightarrow0}\left[2\pi i\frac{C}{2\pi}\left(\frac{e^{i\left(\omega_{2}-i\epsilon\right)\left(t-t'\right)}}{2\omega_{2}} -\frac{e^{-i\left(\omega_{2}+i\epsilon\right)\left(t-t'\right)}}{2\omega_{2}}\right)\right].
\end{equation}
After taking the $\epsilon\rightarrow 0$ limit, we have
\begin{equation}
F(t-t')=-\frac{C}{\omega_{2}}\sin\left[\omega_{2}\left(t-t'\right)\right].
\end{equation}
Substituting this result back into Eq. (\ref{eq: IntOut})
\begin{equation}
\left(\phi_{2}\right)_{ret}(t)=\int dt'\left(-\frac{C}{\omega_{2}}\right)\theta\left(t-t'\right) \sin\left[\omega_{2}\left(t-t'\right)\right]\phi_{1}(t'),
\end{equation}
where $\theta\left(t-t'\right)$ is the Heaviside function, which
enforces causality.

\section{Background STF tensors
\label{subsec:3(B)-Background-quadrupolar}}

Using the definition of the background STF tensor in Eq.~(\ref{eq:BGquadrupolarSTF_def})
and the proper time dependence of the vector $n$, we can express
$Q^{BG}_{ij}$ as a 3 dimensional spatial matrix:
\begin{equation}
Q^{BG}_{ij}=\frac{\mu}{r^{3}}\left(\begin{array}{ccc}
3\cos^{2}\left(\Omega\tau\right)-1 & \frac{3}{2}\sin\left(2\Omega\tau\right) & 0\\
\frac{3}{2}\sin\left(2\Omega\tau\right) & 3\sin^{2}\left(\Omega\tau\right)-1 & 0\\
0 & 0 & -1
\end{array}\right),
\end{equation}
where $\Omega$ is the orbital frequency, $\mu=\frac{q}{1+q}M$ is
the reduced mass, $q<1$ is the dimensionless mass ratio of the binary,
$M$ is the heavier companion mass, and $r$ is the orbital separation.
When we contract $Q^{BG}(\tau)$ with $Q^{BG}(\tau')$, we obtain
\begin{equation}
Q^{BG}_{ij}(\tau)Q^{BG\ ij}(\tau') =9\frac{\mu^{2}}{r^{6}} \left[\cos^{2}\left[\Omega\left(\tau-\tau'\right)\right]-\frac{1}{3}\right],
\end{equation}
note that we have treated the orbital separation as time-independent, but since it varies slowly throughout most of the inspiral phase, we neglect
the small change in $r$. For the instanteneous contraction, $\tau'=\tau$,
\begin{equation}
Q^{BG}_{ij}(\tau)Q^{BG\ ij}(\tau)=6\frac{\mu^{2}}{r^{6}},
\label{QBG2}
\end{equation}
which can also be written as
\begin{equation}
Q^{BG}_{ij}(\tau)Q^{BG\ ij}(\tau)=6\frac{q^{2}}{\left(1+q\right)^{2}}\frac{M^{2}}{r^{6}}
\simeq 6~{q^{2}}\frac{M^{2}}{r^{6}}.
\end{equation}

\section{ Retarded response kernel
\label{subsec:(C)-Retarded-response}}

To enforce causality, we define the retarded response kernel,
\begin{equation}
K\left(\tau,\tau'\right)=\omega_{2}\sin\left[\omega_{2}\left(\tau-\tau'\right)\right]\theta(\tau-\tau'),
\end{equation}
so that, without making a low frequency approximation, (\ref{eq:induced quadrupole})
can be written as
\begin{equation}
Q^{I}_{ij}(\tau)=\int d\tau'\ K\left(\tau,\tau'\right)Q^{BG}_{kl}(\tau').
\end{equation}
This introduces integrals of the form
\begin{equation}
I_{B}=\int d\tau\ \int d\tau'\ \omega_{2}\sin\left[\omega_{2}\left(\tau-\tau'\right)\right]\theta(\tau-\tau')F(\tau-\tau'),
\end{equation}
with some function $F(\tau-\tau').$ In such cases, a useful identity
is
\begin{equation}
\omega_{2}\sin\left[\omega_{2}\left(\tau-\tau'\right)\right]=\frac{\partial}{\partial\tau'}\cos\left[\omega_{2}\left(\tau-\tau'\right)\right],
\end{equation}
which is convenient when performing integration by parts with respect
to $\tau'$. Dropping the boundary terms at $\tau'=\pm\infty$,
the integral becomes the following
\be
I_{B}=\int d\tau\ \int d\tau'\ \left[\delta(\tau-\tau')F(\tau-\tau') -\theta(\tau-\tau') \frac{\partial}{\partial\tau'}F(\tau-\tau')\right]\cos\left[\omega_{2}\left(\tau-\tau'\right)\right].
\ee

\section{Stationary Phase Approximation
\label{subsec:(D)-Resonance-condition}}

Here, we employ a modified stationary-phase approximation that relies on the same underlying assumption as the conventional method, namely that rapidly oscillating contributions to the integral average to zero, but differs in the prescription used to isolate and evaluate the dominant contribution.
To solve an integral of the form
\begin{equation}
I_{C}=\int^{\tau}_{-\infty}d\tau'\ \Omega\left[\sin\left[\left(2\Omega+\omega_{2}\right)\left(\tau-\tau'\right)\right] +\sin\left[\left(2\Omega-\omega_{2}\right)\left(\tau-\tau'\right)\right]\right],
\end{equation}
we rewrite $I_C$ as,
\begin{equation}
I_{C}=Im\left[\int^{\tau}_{-\infty}d\tau'\ \Omega\left(e^{i\left(2\Omega+\omega_{2}\right)\left(\tau-\tau'\right)} +e^{i\left(2\Omega-\omega_{2}\right)\left(\tau-\tau'\right)}\right)\right].
\end{equation}

At a generic instant during the inspiral phase, the orbital frequency varies slowly while both exponential terms oscillate rapidly and therefore do not contribute significantly to the integral.
However, near the resonance time
$\tau'=\tau_{R}$,
which is defined as the time at which the resonance condition holds
$
2\Omega(\tau'=\tau_{R})=\omega_{2}
$,
the integral does receive a significant contribution.
We restrict our attention to the case that the orbital frequency has the same sign as that of $\omega_{2}$ \footnote{The resonance condition can also be satisfied as $2\Omega(\tau'=\tau_{R})=-\omega_{2}$, and leads to similar results.}.

The resonance contribution dominates the integral, and we therefore neglect the nonresonant term.
We expand the phase of the resonance term to second order
\begin{equation}
\left(2\Omega-\omega_{2}\right)\left(\tau-\tau'\right)\simeq  2\dot\Omega_R (\tau-\tau_R)(\tau'-\tau_R)+\frac{1}{2}\left(2\ddot\Omega_R(\tau-\tau_R)-2\dot\Omega_R\right)(\tau'-\tau_R)^2,
\end{equation}
where
\begin{equation}
\dot{\Omega}_{R}=\frac{\partial}{\partial\tau'}\Omega\bigg|_{\tau'=\tau_{R}},\quad\ddot{\Omega}_{R}=\frac{\partial^2}{\partial\tau'^2}\Omega\bigg|_{\tau'=\tau_{R}}.
\end{equation}
Since the orbital frequency grows monotonically during the inspiral, $\dot{\Omega}_{R}>0$.

The second time derivative can be neglected relative to the first derivative, as we explain in Appendix \ref{subsec:(E)-Omega-Dot}. We expand the orbital frequency around the resonance time,
\begin{equation}
\Omega\simeq\frac{\omega_{2}}{2}+\dot{\Omega}_{R}\left(\tau'-\tau_{R}\right),
\end{equation}
after rearranging the argument of the exponential, we obtain
\bea
I_{C}&\simeq&\theta\left(\tau-\tau_{R}\right)~Im\bigg[\int^{\tau}_{-\infty}d\tau'\ \left(\frac{\omega_{2}}{2}+\dot{\Omega}_{R}\left(\tau'-\tau_{R}\right)\right) \times
\cr &&
 \exp\left(-i2\dot{\Omega}_{R}\left(\tau'-\frac{\tau+\tau_{R}}{2}\right)^{2}\right) \exp\left(i\frac{\dot{\Omega}_{R}}{2}\left(\tau-\tau_{R}\right)^{2}\right)\bigg],
\eea
where we have included the Heaviside function $\theta(\tau-\tau_{R})$
to ensure that the resonance time lies within the integration interval.
The result is
\begin{equation}
I_{C}=\sqrt{\frac{\pi}{8\dot{\Omega}_{R}}}\theta\left(\tau-\tau_{R}\right) \left(\omega_{2}+\dot{\Omega}_{R}\left(\tau-\tau_{R}\right)\right)\sin\left[\frac{\dot{\Omega}_{R}}{2} \left(\tau-\tau_{R}\right)^{2}-\frac{\pi}{4}\right].
\end{equation}

In a similar manner we can compute the integral
\begin{equation}
I_{D}=\int^{\tau}_{-\infty}d\tau'\ \Omega\left[\cos\left[\left(2\Omega+\omega_{2}\right) \left(\tau-\tau'\right)\right] +\cos\left[\left(2\Omega-\omega_{2}\right)\left(\tau-\tau'\right)\right]\right],
\end{equation}
using Eulre's identity,
\begin{equation}
I_{D}=Re\left[\int^{\tau}_{-\infty}d\tau'\ \Omega\left(e^{i\left(2\Omega+\omega_{2}\right)\left(\tau-\tau'\right)} +e^{i\left(2\Omega-\omega_{2}\right)\left(\tau-\tau'\right)}\right)\right].
\end{equation}
Applying the stationary phase approximation we find,
\begin{equation}
I_{D}\simeq\sqrt{\frac{\pi}{8\dot{\Omega}_{R}}}\theta\left(\tau-\tau_{R}\right) \left(\omega_{2}+\dot{\Omega}_{R}\left(\tau-\tau_{R}\right)\right) \cos\left[\frac{\dot{\Omega}_{R}}{2}\left(\tau-\tau_{R}\right)^{2}-\frac{\pi}{4}\right].
\end{equation}

\section{Hierarchy of the time derivatives of the orbital angular frequency
\label{subsec:(E)-Omega-Dot}}

To examine the time dependence of the orbital angular frequency
during the inspiral, we begin with Kepler's third law,
\begin{equation}
    \Omega^2
    = \frac{1}{2}(1+q)
    \left(\frac{R_S}{r}\right)\frac{1}{r^2},
\end{equation}
where $q=m/M\ll 1$ is the binary mass ratio and $R_S=2M$ is the
Schwarzschild radius of the heavier object.
Taking the time derivative gives
\begin{equation}
    \dot{\Omega}
    = -\frac{3}{2}\Omega\frac{\dot{r}}{r}.
\end{equation}
For a quasi-circular binary, the leading-order quadrupole
formula for the orbital decay yields
\begin{equation}
    \dot{r}
    = -\frac{8}{5}q(1+q)
    \left(\frac{R_S}{r}\right)^3.
\label{rdot}
\end{equation}
Substituting this expression into the first derivative
of the orbital frequency, and using Kepler's law to
eliminate $1/r$, we obtain
\begin{equation}
    \frac{\dot{\Omega}}{\Omega}
    = \frac{12\sqrt{2}}{5}
    q\sqrt{1+q}
    \left(\frac{R_S}{r}\right)^{5/2}\Omega.
\label{dotObyO}
\end{equation}

To evaluate the second time derivative, we differentiate
the expression for $\dot{\Omega}$ once more,
\begin{equation}
    \ddot{\Omega}
    = -\frac{3}{2}\Omega
    \left[
        \frac{\ddot{r}}{r}
        - \left(\frac{\dot{r}}{r}\right)^2
    \right]
    -\frac{3}{2}\dot{\Omega}\frac{\dot{r}}{r}.
\end{equation}
The quadrupole formula implies
$\ddot{r}=-3\dot{r}^{\,2}/r$.
Consequently, the second derivative simplifies to
\begin{equation}
    \ddot{\Omega}
    =-\frac{11}{2}\dot{\Omega}\frac{\dot{r}}{r},
\end{equation}
and therefore
\begin{equation}
    \frac{\ddot{\Omega}}{\dot{\Omega}}
    =\frac{11}{2}\frac{\dot{\Omega}}{\Omega}
    =\frac{66\sqrt{2}}{5}
    q\sqrt{1+q}
    \left(\frac{R_S}{r}\right)^{5/2}\Omega.
\label{ddotO}
\end{equation}

Thus, the dimensionless ratios characterizing the
variation of the orbital frequency satisfy
\begin{equation}
    \frac{\dot{\Omega}}{\Omega^2}
    =\frac{12\sqrt{2}}{5}
    q\sqrt{1+q}
    \left(\frac{R_S}{r}\right)^{5/2},
    \qquad
    \frac{\ddot{\Omega}}{\Omega\dot{\Omega}}
    =\frac{11}{2}\frac{\dot{\Omega}}{\Omega^2}.
\end{equation}
Both quantities are small in the slow-inspiral regime,
$q(R_S/r)^{5/2}\ll 1$,
demonstrating that the orbital frequency and its first
time derivative vary slowly compared with the orbital
timescale. This motivates a Taylor expansion of the
orbital frequency around the resonance time $\tau_R$,
defined by $2\Omega(\tau_R)=\omega_2$:
\begin{equation}
    \Omega(\tau)
    =\Omega(\tau_R)
    +\dot{\Omega}(\tau_R)(\tau-\tau_R)
    +\frac{1}{2}\ddot{\Omega}(\tau_R)
        (\tau-\tau_R)^2+\cdots.
\end{equation}

We may estimate the elapsed time from resonance $\tau-\tau_R$ by integrating Eq.~(\ref{rdot}),
\be
\tau-\tau_R = \frac{5}{32}~\frac{1}{q} \frac{r_R^4-r^4}{R_S^3}.
\ee
Thus, using Eq.~(\ref{dotObyO}),
\be
  \frac{\dot\Omega(\tau-\tau_R)}{\Omega}=\frac{3}{8}\left(1-\frac{r}{r_R}\right)
\ee
Similarly, using Eq.~(\ref{ddotO}),
\be
 \frac{1}{2}\frac{\ddot\Omega(\tau-\tau_R)}{\dot\Omega}\simeq 1-\frac{r}{r_R}
\ee
Consequently, for the case that $r_{ISCO}/r_R$ is not parametrically small,  higher time derivatives of $\Omega$ are suppressed with respect to the leading order during the inspiral.

\bibliographystyle{plainurl}

\end{document}